\documentclass[12pt]{iopart}
\usepackage{iopams}  
\usepackage{setstack}
\usepackage{graphics}
\usepackage[OT4]{fontenc}
\begin{document}

\title[Energy losses of fast heavy-ion projectiles in dense hydrogen plasmas]{Energy losses of fast heavy-ion projectiles in dense hydrogen plasmas}
\author{D Ballester$^{1}$ and I M Tkachenko$^{2}$}
\address{$^{1}$ School of Mathematics and Physics, Queen's University, Belfast BT7 1NN, United Kingdom}
\address{$^{2}$ Departamento de Matem\'{a}tica Aplicada, Universidad Polit\'{e}cnica de Valencia, 46022 Valencia, Spain}

\eads{d.berman@qub.ac.uk, imtk@mat.upv.es}
\begin{abstract}
It has been recently shown that the Bethe-Larkin formula for the energy losses of fast heavy-ion projectiles in dense hydrogen plasmas is corrected by the electron-ion correlations [Phys. Rev. Lett. \textbf{101}, 075002 (2008)]. We report numerical estimates of this correction based on the values of $g_{ei}(0)$ obtained by numerical simulations in [Phys. Rev. E \textbf{61}, 3470 (2000)]. We also extend this result to the case of projectiles with dicluster charge distribution. We show that the experimental visibility of the electron-ion correlation correction is enhanced in the case of dicluster projectiles with randomly orientated charge centers. Although we consider here the hydrogen plasmas to make the effect physically more clear, the generalization to multispecies plasmas is straightforward. 
\end{abstract}

\maketitle

\section{Introduction}

The calculation of the stopping power for a projectile traversing a plasma can be performed within the so called polarizational approach. The Lindhard formula relates the polarizational stopping power with the (longitudinal) dielectric function of the medium \cite{Li}. When a projectile with a charge distribution $\rho(\mathbf{r},t)$ moves at a constant velocity $\mathbf{v}$ through an isotropic Coulomb fluid, this formula can be written as \cite{BD93}
\begin{equation}
-    \frac{dE}{dx}  =\frac{1 }{2 \pi^ 2
v }\int {{\rm d} \mathbf{k}} \frac{\mathbf{k}\cdot \mathbf{v}}{k^2} 
\left(- {\rm Im} \epsilon^{-1} \left(
k, \mathbf{k}\cdot \mathbf{v} \right)  \right) | \rho(\mathbf{k}) |^2  , \label{dedxpol1}%
\end{equation}
where $k = | \mathbf{k} |$. The polarizational approach to the stopping power calculation is justified provided that the interaction between the
projectile and the plasma is so weak that it can be considered a linear effect. In fact, one can expect that at projectile velocities much higher than the characteristic velocity of charged particles of the medium (e.g. the Fermi velocity $v_F$), the polarizational mechanism is responsible for the main contribution to the energy losses.

In order to make the calculations analytically tractable, it is usually assumed that the collective plasmon mode constitutes the most important channel of energy transfer from fast external projectiles to the plasma. If we admit the existence of a well-defined plasma mode with negligible damping, we can write \cite{Tka}
\begin{equation}
-\frac{{\rm Im}\epsilon
^{-1}\left(  k,\omega\right)  }{\pi C_{0}\left(  k\right) \omega} 
=    \frac{\omega_{2}^{2}-\omega_{1}^{2}}{\omega_{2}^{2}}\delta\left(
\omega\right)  +\frac{\omega_{1}^{2}}{2\omega_{2}^{2}}\left[  \delta\left(
\omega-\omega_{2}\right)  +\delta\left(  \omega+\omega_{2}\right)  \right]   ,
\label{can}%
\end{equation}
where $\omega_{1}^{2}  =C_{2}/C_{0} $, $\omega_{2}^{2} =C_{4}/C_{2} $, $C_{0}\left(  k\right)$, $C_{2}\left(  k\right)$ and $C_{4}\left(  k\right)$ are the frequency moments of the plasma loss function representing the zero-frequency sum rule, the f-sum rule and the third-frequency-moment sum rule, respectively, in completely ionized plasmas \cite{kugler,amt,mt,AT}.

\section{Point-like projectiles}

For a fast projectile with a point-like distribution the stopping power formula (\ref{dedxpol1}) can be evaluated using expression (\ref{can}) to give \cite{BT08}, in a completely ionized hydrogen plasma,
\begin{equation}
\left( - \frac{dE}{dx} \right)_{cBL} \underset{v \gg v_{F} }{\simeq
} \left(  \frac{Z_{p}e\omega_{p}}{v}\right)  ^{2}\ln\frac{2m v^{2}  } {\hbar\omega_{p}\sqrt{1+ H }} ,\label{modBL}%
\end{equation}
\begin{equation}
H=  \frac{  g_{ei}\left(
0\right)  -1 } { 3} =\frac{1}{6\pi^{2}n }%
\int _{0}^{\infty}
p^{2}S_{ei}\left(  p\right)  {\rm d}p , \label{H}
\end{equation}
where $Z_{p}e$ is the projectile charge, $\omega_{p}=\left( 4\pi n e^{2}/m \right)^{1/2}$ is the plasma frequency,
$n$ is the electron number density, and $m$ is the electron mass. In addition, $g_{ei}(0)$ is the value of the electron-ion radial distribution function at the origin and $S_{ei}(k)$ is the electron-ion partial static structure factor. This result implies that the well-known Bethe-Larkin (BL) formula \cite{bethe} for the stopping power gets corrected due to the electron-ion correlation in the plasma (through the quantity $H$). In case of a one-component plasma, the existence of electron-electron correlations does not affect the fast projectile stopping power \cite{OT01}.

While the correction $H$ (or $h_{ei}(0)=g_{ei}(0)-1$) vanishes in a weakly coupled plasma, when dealing with plasmas within the strong coupling regime $H$ might become significant. In Table \ref{table} the values of $g_{ei}(0)$ obtained by numerical simulations in Ref. \cite{MP} are displayed. The simulated system was an electron-proton plasma at 125000K and with different values of the Brueckner parameter, $r_s$. In the third column of Table \ref{table} we quantify, by means of the parameter
\begin{equation}
\Delta=1-\frac{(-dE/dx)_{cBL}}{(-dE/dx)_{BL}} = \frac{\ln \sqrt{1+H}}{\ln \frac{2mv^2}{\hbar \omega_p}}, \label{delta}
\end{equation}
the effect the correction $H$ has over the stopping power asymptotic value given in (\ref{modBL}), in this way we compare $(-dE/dx)_{cBL}$ to the BL formula, $(-dE/dx)_{BL}$, i.e., (\ref{modBL}) but with $H\equiv 0$. In addition, for the numerical estimates the projectiles were assumed to be protons at 1 MeV. While the results indicate that the Bethe-Larkin formula gets modified by some $7\div 11\%$, in a more realistic scenario one should account for other effects that could limit the experimental visibility of this correction. One of those factors is the existence of damping of the plasma mode, which would lead to the broadening of the plasmon peak and could modify the value of the stopping power hiding the correlation shift of the Langmuir mode \cite{BT08}. Furthermore, the experimental techniques available at the present time only allow to measure the stopping power with a significant inaccuracy \cite{X}. Thus, it is desirable to derive alternative methods of diagnostic where the correlation correction $H$ to the stopping power could give rise to larger departures from its uncorrelated counterpart.

\begin{table}
\begin{center}
\begin{tabular}{|c c c|}\hline
 $r_s$ &  $g_{ei}(0)$  &  $\Delta$  \\ \hline\hline $1.75$ & $4.0$ & $7.4\%$ \\ \hline $2$ & $5.0$ & $8.7\%$ \\
\hline $3$ & $7.5$ & $10.5\%$ \\ \hline $4$ & $10.0$ & $11.7\%$ \\ \hline
\end{tabular}
\end{center}
\caption{Modification of the Bethe-Larkin formula due to the electron-ion correlations. The values of $g_{ei}(0)$ correspond to the electron-proton plasma at $125000\rm K$, as simulated in Ref. \cite{MP}. The projectiles are assumed to be protons at $1$ $\rm MeV$.} \label{table}
\end{table}

\section{Dicluster projectiles}

Let us consider then the stopping of projectiles consisting of diclusters with (in general, unequal) charges $Z_1$, $Z_2$ whose spatial distribution (at rest) is given by $\rho(\mathbf{r}) = Z_1 e \delta(\mathbf{r}) + Z_2 e \delta(\mathbf{r} - \mathbf{R})$ \cite{BD93}. Under the assumption that both charge centers of the dicluster are randomly orientated, we can write:
\begin{equation}
|\rho(\mathbf{k })|^2  = (Z_1 e)^2 +  (Z_2 e)^2 + 2  Z_1 Z_2 e^2  j_0 (Rk)  , \label{distdicl}%
\end{equation}
where $j_0(x)$ is the zero-order spherical Bessel function. After substituting this charge distribution into Eq. (\ref{dedxpol1}) one gets \cite{BD93}

\begin{equation}
- \frac{dE}{dx} = \left(- \frac{dE}{dx}\right)_{uncorr} +\left(- \frac{dE}{dx}\right)_{corr} ,\label{stopdicl}%
\end{equation}
where the first term represents the stopping power of uncorrelated charge centers. It simply reduces to the same expression valid for a point-like distribution, but with the charge number $\sqrt{ Z_1^2 + Z_2^2 }$ instead of $Z_p$. The second contribution, on the other hand, describes the interference pattern stemming from the correlation between the charge centers of the dicluster distribution \cite{BD93}: 
\begin{equation}
\left(- \frac{dE}{dx}\right)_{corr} = \frac{2e^2}{\pi v^2} 2  Z_1 Z_2   \int_0^\infty  j_0 (Rk)  \frac{{\rm d}k}{k} \int_0^{kv} \omega 
 \left(- {\rm Im} \epsilon^{-1} \left(
k, \omega \right)  \right) {\rm d}\omega  .   \label{diclcorr}%
\end{equation}
Obviously, this expression vanishes as the distance between the charge centers, $R$, increases. When $R\to 0$, the whole expression (\ref{stopdicl}) reduces to the stopping power of a point-like distribution with the charge number $(Z_1 + Z_2)$.

\begin{figure}
\centering{
\includegraphics{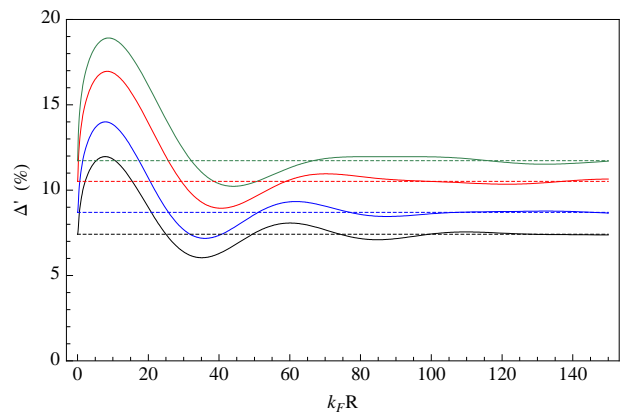}
\caption{\label{figure} Modification of the uncorrelated dicluster stopping power with $Z_1=Z_2$ due to the electron-ion correlation as a function of the distance between the charge centers of the dicluster. The same four conditions used in Table \ref{table} are considered: $r_s =1.75$ (bottom curve), $r_s =2$, $r_s =3$ and $r_s =4$ (top curve).} } 
\end{figure}

We can derive the fast-projectile asymptote for the correlated contribution to the dicluster stopping by substituting expression (\ref{can}) into the previous formula. After applying the arguments presented in Ref. \cite{BT08}, we obtain that
\begin{equation}
\left(- \frac{dE}{dx}\right)_{corr}   \underset{v \gg v_{F} }{\simeq}
 2Z_1 Z_2 \left( \frac{e\omega_{p}}{v}\right)  ^{2}  \left[ \mathcal{H}(k_{max}R) -  \mathcal{H}(k'_{min}R)\right] ,\label{modcorr}%
\end{equation}
with $ \mathcal{H}(x) = {\rm Ci}(x) - j_0 (x)$, ${\rm Ci}(x)$ being the cosine integral \cite{BD93}, $k_{max}=2mv/\hbar$ and $k'_{min}=\omega_{p}\sqrt{1+H}/v$.

As in the case of point-like projectiles in (\ref{delta}), the difference between the uncorrected dicluster stopping power ($H\equiv 0$) and the corrected one for $Z_{1}=Z_{2}$ is given by
\begin{equation}
\Delta'= \frac{\ln \sqrt{1+H}+ \left[ \mathcal{H}(k'_{min}R) - \mathcal{H}(k_{min}R) \right] }{\ln \frac{2mv^2}{\hbar \omega_p} + \left[ \mathcal{H}(k_{max}R) - \mathcal{H}(k_{min}R) \right]  }, \label{deltaprime}
\end{equation}
with $k_{min}=\omega_p /v$. This quantity is plotted in Fig. \ref{figure}, depicting its variation as a function of the distance between the dicluster charge centers for the thermodynamic conditions considered in Table \ref{table}. Whereas the value of the parameter $\Delta'$ approaches that of $\Delta$ as $R\to 0$ and $R\to \infty$, the interference between the charge centers of the dicluster produces an oscillatory pattern for $\Delta'$ as $R$ varies, showing that it is feasible to find numerical intervals for $R$ where the dicluster stopping power exhibits a larger departure from the uncorrelated stopping than the one observed for the point-like stopping power discussed previously. For instance, in the case $r_s =4$ the value of $\Delta'$ increases from an $11.7\%$ to a $19\%$, for the optimal value of $R$. 

Although an extensive experimental work will be required to understand their practical applicability, the methods considered here might become suitable candidates as diagnostic tools in high energy density plasmas, where other techniques fail \cite{X}, not only to determine the charge number density, but even to retrieve the value of $g_{ei}(0)$, providing a method of experimental investigation of strong coupling effects in Coulomb systems \cite{BT08}.

\section*{Acknowledgements}

D.B. is grateful to C. Di Franco for discussions and insights. The authors acknowledge the financial support of the European Social Fund, the Spanish Ministerio de Educaci\'on y Ciencia (Project No. ENE2007-67406-C02-02/FTN), and the INTAS (Project No. 06-1000012-8707).

\end{document}